# Direct measurement of the three-dimensional magnetization vector trajectory in GaMnAs by a magneto-optical pump-and-probe method

N. Tesařová, P. Němec[a)], E. Rozkotová, J. Šubrt, H. Reichlová,
D. Butkovičová, F. Trojánek, and P. Malý
*Faculty of Mathematics and Physics, Charles University in Prague, Ke Karlovu 3,
121 16 Prague 2, Czech Republic*

V. Novák
*Institute of Physics ASCR v.v.i., Cukrovarnická 10, 162 53 Prague, Czech Republic*

T. Jungwirth
*Institute of Physics ASCR v.v.i., Cukrovarnická 10, 162 53 Prague, Czech Republic
and School of Physics and Astronomy, University of Nottingham, Nottingham NG7 2RD,
United Kingdom*

We report on a quantitative experimental determination of the three-dimensional magnetization vector trajectory in GaMnAs by means of the static and time-resolved pump-and-probe magneto-optical measurements. The experiments are performed in a normal incidence geometry and the time evolution of the magnetization vector is obtained without any numerical modeling of magnetization dynamics. Our experimental method utilizes different polarization dependences of the polar Kerr effect and magnetic linear dichroism to disentangle the pump-induced out-of-plane and in-plane motions of magnetization, respectively. We demonstrate that the method is sensitive enough to allow for the determination of small angle excitations of the magnetization in GaMnAs. The method is readily applicable to other magnetic materials with sufficiently strong circular and linear magneto-optical effects.

The magnetic data storage relies on setting the magnetization orientation along a certain direction in a ferromagnetic material. The speed of data storage is connected with the dynamical response of magnetization to external stimuli. A direct experimental determination of the time-dependent non-equilibrium magnetization vector is therefore desirable because of the basic understanding of magnetization dynamics as well as for practical applications of the magnetization switching phenomena. In the last decade, several variants of experimental stroboscopic magneto-optical (MO) methods have been reported that enable to measure the real-time trajectory of non-equilibrium magnetization [1-5]. In particular, the out-of-plane component of magnetization is well accessible due to the polar magneto-optical Kerr effect (MOKE). Here a linearly polarized light experiences a change of polarization after the reflection on a magnetized medium and the magnitude of this polarization change is proportional to the projection of the magnetization along the light propagation. The availability of femtosecond lasers together with a relative simplicity of the corresponding experimental setup made a time-resolved MOKE the most effective experimental tool for the measurement of the ultrafast magnetization dynamics [6]. However, in a typical experimental setup – with a rather small angle between the light propagation direction and the normal to the sample surface (angle of incidence, $\theta_i$, in the following), only the out-of-plane component of the magnetization is detected [7]. To measure the in-plane components of the time dependent

---

magnetization, it is necessary to employ another magneto-optical phenomenon than the polar MOKE. The most common techniques utilize the change of light polarization due to the longitudinal MOKE [2, 3, 8], the change of light intensity due to transversal MOKE [1], the interpretation of reflected light intensity at different polarizations by the Fresnel matrix formalism [5], or the second-harmonic MOKE technique [4] (see the Supplementary material for a detailed discussion of the corresponding experimental methods and their limitations [9]). With these methods, it is however a significant experimental challenge to perform a quantitative, high-sensitivity measurement of the time dependence of the full three-dimensional magnetization vector [9]. In this paper we introduce an experimental technique which utilizes a different MO effect – the magnetic linear dichroism – for measuring of the in-plane component of the time dependent magnetization. Importantly, this MO effect can be observed in exactly the same sample orientation and experimental setup as the polar MOKE which enables to perform direct quantitative measurements of the full three-dimensional magnetization vector evolutions.

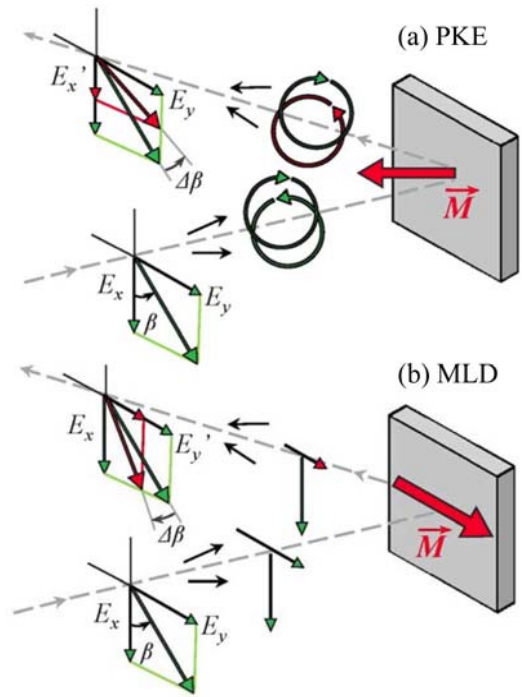

Fig. 1. Schematic illustration of two magneto-optical effects in (Ga,Mn)As that are responsible for a rotation of the polarization plane $\Delta\beta$ of reflected light at normal incidence. (a) Polar Kerr effect (PKE) that is due to the different index of refraction for $\sigma^+$ and $\sigma^-$ circularly polarized light propagating parallel to the direction of magnetization **M**. (b) Magnetic linear dichroism (MLD) that is due to the different absorption (reflection) coefficient for light linearly polarized parallel and perpendicular to **M** if the light propagates perpendicular to the direction of **M**. $E_x$ and $E_y$ are the projections of the light amplitude to the crystallographic directions [100] and [010], respectively.

Diluted ferromagnetic semiconductors, with (Ga,Mn)As as the most thoroughly investigated example, are magnetic materials that are prepared by a partial replacement of non-magnetic atoms by the magnetic ones [10, 11]. Even thought the achieved Curie temperature is still well below room temperature [10, 11], their research can provide fundamental insight into new physical phenomena that are present also in other types of magnetic materials – like ferromagnetic metals – where they can be exploited in realistic spintronic applications [10-12]. In 2005, a giant magnetic linear dichroism (MLD) was reported in (Ga,Mn)As [13]. As pointed out above, the outstanding feature of this MO effect is that it provides an access to the in-plane component of the magnetization even at normal incidence and that it can be unambiguously separated from the polar Kerr effect (PKE) by its dependence on the orientation of probe beam polarization plane [9, 13]. In Fig. 1 we schematically illustrate the origin of PKE and MLD. In PKE the rotation of light polarization (or the change of its ellipticity) occurs due to the different index of refraction for $\sigma^+$ and $\sigma^-$ circularly polarized light propagating parallel to the direction of magnetization **M**. This polarization rotation depends linearly on the magnitude of the out-of-plane component of **M** and is independ of the orientation of the light polarization plane $\beta$ [9]. On the other hand, MLD originates from the different absorption (reflection) coefficient for the light polarization plane oriented parallel and perpendicular to **M** [9]. This effect scales quadratically with the in-plane magnetization component and varies as $\sin(2\beta)$ [9].

We investigated the laser-pulse induced dynamics of magnetization by the well known pump-and-probe MO technique where the output of a femtosecond laser is divided into a strong pump pulse and a weak probe pulse that are focused to a same spot on the sample [6, 7]. Laser pulses, with the time width of 200 fs and the repetition rate of 82 MHz, were tuned to 1.64 eV (i.e., above the semiconductor band gap). The fluence of the pump pulses is 30 $\mu$J.cm$^{-2}$, which corresponds to the photoinjected carrier density of about $1.7 \times 10^{18}$ cm$^{-3}$, and probe pulses were twenty times weaker. The experiment was performed close to the normal incidence ($\theta_i = 2°$ and 8° for pump and probe pulses, respectively) with a sample placed in a cryostat and cooled down to the temperature of about 15 K. The time-resolved data reported here were obtained without any external magnetic field applied. However, prior to this time-resolved experiment, the magnetization was oriented in a well defined position (so-called easy axis) by an application of 500 mT along the [010] crystallographic direction in the sample plane (see Fig. 1 and its figure caption for a definition of the coordinate system). The dynamical MO data shown here correspond to the pump-hellicity-independent part of the measured signals [14]. We also confirmed that the measured dynamical MO signal reflects the magnetization dynamics by comparing the signal corresponding to the probe rotation and elipticity change [15]. The orientation of polarization plane of linearly polarized probe pulses $\beta$, which is measured from the [100] crystallographic direction, was changed by a wave plate. The magnitudes of MO coefficients of PKE ($P^{PKE}$) and MLD ($P^{MLD}$) were measured in separate static magneto-optical experiments where an external magnetic field of 500 mT was used to align the magnetization in the out-of-plane and in-plane orientations, respectively. All the reported experiments were performed in several samples with a different Mn content from our set of high-quality epilayers which are as close as possible to uniform uncompensated Ga$_{1-x}$Mn$_x$As mixed crystals [16]. The obtained results are rather similar for all the samples and, therefore, we report here only results measured in a sample with nominal doping $x = 5.2\%$ and Curie temperature $T_c = 132$ K. Magnetic anisotropy constants and easy-axis orientation were independently determined by SQUID magnetization measurements [16].

In Fig. 2(a) we show measured time-resolved MO signals. The observed oscillatory signal is a signature that an impact of the pump pulse induces a precession of magnetization in the sample [17-20]. Recall that no external magnetic field was applied during this experiment and, therefore, the precession frequency is solely given by the internal magnetic anisotropy fields. The most striking feature apparent from Fig. 2(a) is that – at identical pumping conditions – the measured dynamical MO data are strongly dependent on the probe polarization orientation $\beta$. This behavior was already reported in Ref. 18 where it was assigned to the contribution of MLD to the measured MO signal. However, as the MO coefficients of the sample used in Ref. 18 were not known to the authors, the numerical analysis of the data by the Landau-Lifshitz-Gilbert (LLG) equation was only qualitative [18].

In Fig. 2(b) we show the spectral dependence of magneto-optical coefficients, obtained from the static magneto-optical measurements, describing the magnitude of PKE and MLD in the studied sample. It reveals that at the spectral position where the experiment was performed, MLD is merely five-times smaller than PKE which explains the strong dependence of the dynamical MO signal on $\beta$ [see Fig. 2(a)]. To disentangle the various contributions in the measured time dependent data it is quite illustrative to perform the following analysis. The measured dynamical MO signal $\delta MO$, which is a function of the time delay between pump and probe pulses $\Delta t$ and of $\beta$, can be fitted well by the phenomenological equation,

$$\delta MO(\Delta t, \beta) = A(\beta)\cos[\omega_{Mn}\Delta t + \Phi(\beta)]e^{-\Delta t/\tau_G} + C(\beta)e^{-\Delta t/\tau_p}, \qquad (1)$$

where $A$ and $C$ are the amplitudes of the oscillatory and pulse function, respectively, $\omega_{Mn}$ is the ferromagnetic moment precession frequency, $\Phi$ is the phase factor, $\tau_G$ is the Gilbert damping time, and $\tau_p$ is the pulse function decay time. All the measured data in Fig. 2(a) can be fitted reasonably well by Eq. (1) with a one set of parameters $\omega_{Mn}$, $\tau_G$ and $\tau_p$. The dependences $A(\beta)$ and $C(\beta)$ obtained by this fitting procedure are displayed in Fig. 2(c) and Fig. 2(d), respectively. The magnetization orientation can be characterized by polar ($\varphi$) and azimuthal ($\theta$) angles - see inset in Fig. 3(a) for their definition. Before an impact of the pump pulse the magnetization points along the easy axis direction (characterized by angles $\varphi_0$ and $\theta_0$), which is determined by the magnetic anisotropy of the sample. From the SQUID measurement we know that $\theta_0$ = 90° but the precise value of $\varphi_0$ is difficult to obtain from these data [16]. Absorption of the pump laser pulse leads to a photo-injection of electron-hole pairs and to a transient increase of the lattice temperature [17-19]. This in turn leads to a quasi-equilibrium change of the easy axis position with maximal in-plane and out-of-plane tilts $\delta\varphi_{qe}$ and $\delta\theta_{qe}$, respectively. Consequently, the magnetization starts to precess around the new, quasi-equilibrium easy axis position. The measured oscillatory MO signal contains a component due to the out-of-plane motion of the magnetization, which is sensed by PKE, and a signal due to the in-plane movement of magnetization, which is sensed by MLD. The first one (with an amplitude $A^{PKE}$) does not depend on $\beta$ but the second one (with an amplitude $A^{MLD}$) is a harmonic function of $\beta$ [9]. Due to the precessional motion of magnetization these signals are phase shifted by 90° and the total amplitude of the oscillatory MO signal $A(\beta)$ is given by [9]

$$A(\beta) = \sqrt{\left[\delta\varphi_{qe} P^{MLD} 2 \cos 2(\varphi_0 - \beta)\right]^2 + \left[A^{PKE}\right]^2} \:. \tag{2}$$

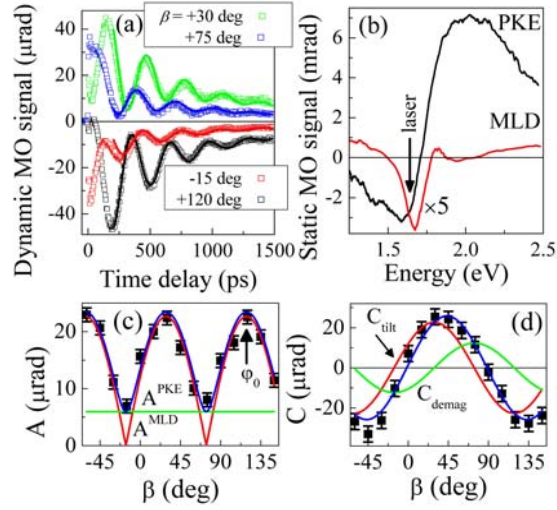

Fig. 2. Magneto-optical signals (polarization rotations) measured in (Ga,Mn)As. (a) Dynamics of the MO signal induced by an impact of pump pulse on the sample that was measured by probe pulses with different polarization orientations $\beta$ (points); lines are fits by Eq. (1) with parameters $\omega_{Mn}$ = 20.2 GHz, $\tau_G$ = 360 ps and $\tau_p$ = 1050 ps. (b) Spectral dependence of static PKE and MLD, the arrow indicate the spectral position of the laser pulses used in the time-resolved experiment shown in part (a); note that the data for MLD are multiplied by 5 for clarity. (c) and (d) Polarization dependence of the oscillation amplitude $A$ (c) and of the pulse function amplitude $C$ (d) at time delay of 200 ps that was obtained by fitting the dynamics shown in (a) (points). Lines are results of simultaneous fits of $A(\beta)$ by Eq. (2) and $C(\beta)$ by Eq. (3) with parameters: $\delta\varphi_{qe}$ = +1.1°, $\delta\theta_{qe}$ = 0°, $\delta M_s/M_0$ = −1% [9]. The deduced position of the easy axis in the sample without the pump pulse $\varphi_0$ = 119° is depicted by the vertical arrow in (c).

Eq. (2) can be used to fit the measured dependence of $A(\beta)$ that enables to deduce the equilibrium position of the easy axis in the sample plane $\varphi_0 = 119 \pm 2°$ [see Fig. 2(c)]. We recall that prior to this experiment we oriented the magnetization along the easy axis that is the closest to the [010] crystallographic direction, which corresponds to $\beta = 90°$.

The obtained detailed understanding of the measured MO signals enables us to perform the full quantitative reconstruction of the real-time trajectory of magnetization from the measured dynamical MO signals *without any numerical modeling*. The measured dynamical MO signal can be expressed as [9]

$$\delta MO(\Delta t, \beta) = -\delta\theta(\Delta t) P^{PKE} + \delta\varphi(\Delta t) P^{MLD} 2 \cos 2(\varphi_0 - \beta) + \frac{\delta M_s(\Delta t)}{M_0} P^{MLD} 2 \sin 2(\varphi_0 - \beta), \tag{3}$$

where the first two terms in Eq. (3) describe the influence of magnetization movement perpendicular to the sample plane and in the sample plane, respectively; $\delta\varphi(\Delta t)$ and $\delta\theta(\Delta t)$ are the corresponding transient tilts relative to the equilibrium values $\varphi_0$ and $\theta_0$. The last term in Eq. (3) describes the MO signal change due to the pump-induced demagnetization [21]. The pulse function in $\delta MO$ signal is a transient non-oscillatory change of the MO signal [9]. Therefore, Eq. (3) can be used to fit the experimentally observed dependence $C(\beta)$ if the functions $\delta\varphi(\Delta t)$ and $\delta\theta(\Delta t)$ are replaced by the corresponding quasi-equilibrium tilts of the easy axis $\delta\theta_{qe}$ and $\delta\varphi_{qe}$ along which the magnetization precesses [9]. In fact, the analysis of $C(\beta)$ is of fundamental importance for the interpretation of the measured MO signals because it enables to determine experimentally whether the precession of magnetization is triggered by the out-of-plane or by the in-plane movement of the easy axis. If the out-of-plane movement were dominant in $C_{tilt}$ [i.e., in the first two terms in Eq. (3)], it would not depend on $\beta$. On contrary, the observed harmonic dependence of $C_{tilt}$ on $\beta$ [see Fig. 2(d)] clearly shows that in the investigated sample the easy axis is tilted in the in-plane direction. Moreover, also the demagnetization contribution to the measured pulse function in $\delta MO$ signal ($C_{demag}$) can be separated by this fitting procedure [see Fig. 2(d)].

The actual orientation of the magnetization at any time delay $\Delta t$ is given by $\varphi(\Delta t) = \varphi_0 + \delta\varphi(\Delta t)$, $\theta(\Delta t) = \theta_0 + \delta\theta(\Delta t)$ and its magnitude is given by $M_s(\Delta t) = M_0 + \delta M_s(\Delta t)$. It is apparent from Eq. (3) [see also Fig. 2(d)] that for $\beta$ equal to $\varphi_0$ and $\varphi_0 - 90°$ the demagnetization does not contribute to the measured dynamical MO signal. Moreover, contributions to $\delta MO$ due to $\delta\varphi(t)$ are maximal and exactly opposite for $\beta = \varphi_0$ and $\beta = \varphi_0 - 90°$. Consequently,

$$\delta\varphi(\Delta t) = \left[\delta MO(\Delta t, \varphi_0) - \delta MO(\Delta t, \varphi_0 - 90°)\right] / \left(4 P^{MLD}\right). \tag{4}$$

Similarly, for $\beta$ equal to $\varphi_0 - 45°$ and $\varphi_0 - 135°$ the contributions due to $\delta\varphi(t)$ are not present in $\delta MO$ and the signal due to the demagnetization is exactly opposite for these two angles that leads to

$$\delta\theta(\Delta t) = -[\delta MO(\Delta t, \varphi_0 - 45°) + \delta MO(\Delta t, \varphi_0 - 135°)]/(2P^{PKE}), \quad (5)$$

$$\frac{\delta M_s(\Delta t)}{M_0} = [\delta MO(\Delta t, \varphi_0 - 45°) - \delta MO(\Delta t, \varphi_0 - 135°)]/(4P^{MLD}). \quad (6)$$

In Fig. 3 we show the dynamics of magnetization that was deduced from the data depicted in Fig. 2(a). We note that the time-evolution of the magnetization determined by this direct experimental procedure is very similar to the one that we obtained by a numerical fitting of the measured data by LLG equation [9]. Moreover, the experimental procedure described above enables to deduce directly also the demagnetization dynamics [see Fig. 3(a)] which is not involved in LLG equation [9].

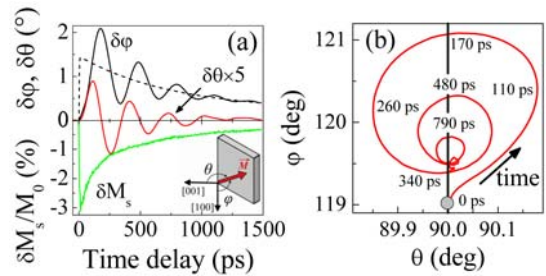

Fig. 3. Direct experimental reconstruction of the magnetization real-space trajectory. An impact of the pump pulse induces a change of the magnetization magnitude and orientation relative to the corresponding equilibrium values (see inset in part (a) for a definition of the polar, $\varphi$, and azimuthal, $\theta$, angles). (a) Time evolution of $\delta\varphi(t)$, $\delta\theta(t)$ and $\delta M_s(t)/M_0$; the dotted line depicts the in-plane evolution of the easy axis position around which the magnetization precesses. (b) Orientation of magnetization at different times after the impact of the pump pulse; the sample plane is represented by the vertical line and the equilibrium position of the easy axis is depicted by the grey spot.

Finally we point out that the applicability of our technique is not limited to the ferromagnetic semiconductor GaMnAs. As the heart of the technique is the simultaneous presence of PKE and MLD in one material, it should be possible to use it in any magnetic material where these two or other similar MO effects (like Faraday and Cotton-Mouton MO effects [22, 23]) are comparable in a certain spectral range. This suggests that other promising candidates for implementation of this technique are, for example, $FeBO_3$ [22] and $DyFeO_3$ [23].

In conclusion, we have demonstrated a magneto-optical, normal incidence pump-and-probe method that enables to perform a quantitative measurement of the real-time trajectory of the full three-dimensional magnetization vector without a need to change the sample position or detection geometry. This method is well suited for detecting small angle magnetization excitations in ferromagnetic semiconductors or other magnetic materials with sufficiently strong circular and linear magneto-optical effects.

This work was supported by the Grant Agency of the Czech Republic grant no. P204/12/0853 and 202/09/H041, by the Grant Agency of Charles University in Prague grant no. 443011 and SVV-2011-263306, by EU grant ERC Advanced Grant 268066 - 0MSPIN, and by Preamium Academiae of the Academy of Sciences of the Czech Republic.

# Direct measurement of the three-dimensional magnetization vector trajectory in GaMnAs by a magneto-optical pump-and-probe method: Supplementary material

N. Tesařová, P. Němec[a)], E. Rozkotová, J. Šubrt, H. Reichlová,
D. Butkovičová, F. Trojánek, and P. Malý
*Faculty of Mathematics and Physics, Charles University in Prague, Ke Karlovu 3,
121 16 Prague 2, Czech Republic*

V. Novák
*Institute of Physics ASCR v.v.i., Cukrovarnická 10, 162 53 Prague, Czech Republic*

T. Jungwirth
*Institute of Physics ASCR v.v.i., Cukrovarnická 10, 162 53 Prague, Czech Republic
and School of Physics and Astronomy, University of Nottingham, Nottingham NG7 2RD,
United Kingdom*

## INTRODUCTION

This supplementary material describes the magneto-optical (MO) response of ferromagnetic semiconductor (Ga,Mn)As. The detailed understanding of the measured MO data enabled us to perform a full quantitative reconstruction of the real-space trajectory of the magnetization precessional motion from the measured dynamical MO signals without any numerical modeling. We also review here the other existing MO methods that can be used for a visualization of the 3-D magnetization movement [1-5] and we discuss their experimental limitations. Moreover, we show that the obtained magnetization dynamics is similar to that deduced by a numerical fitting of the measured MO data by Landau-Lifshitz-Gilbert (LLG) equation.

## BRIEF REVIEW OF EXISTING MAGNETO-OPTICAL TECHNIQUES THAT ENABLE VISUALIZATION OF 3-D MAGNETIZATION MOVEMENT

In the last decade, the commercial availability of femtosecond lasers led to utilization of stroboscopic MO methods for an investigation of various magnetic materials [6]. Due to polar magneto-optical Kerr effect (MOKE), a linearly polarized light experiences a change of polarization after the reflection on a magnetized medium and the magnitude of this polarization change is proportional to the projection of magnetization along the light propagation. However, in a typical experimental setup – with a rather small angle between the light propagation direction and the normal to the sample surface (angle of incidence, $\theta_i$, in the following) – only the out-of-plane dynamics of magnetization is detected [7]. To measure the magnetization movement in other directions, it is necessary to use some other "tool" than polar MOKE.

The most common procedure, which is used to measure the in-plane dynamics of magnetization, is to use a large angle of incidence (with a typical value $\theta_i \approx 45°$) when also other MO effects start to play a role in the detected optical signals. If the magnetization is

---

[a)] Electronic mail: nemec@karlov.mff.cuni.cz

oriented in the light plane of incidence, which is defined by the light direction and the normal to the sample surface, the longitudinal MOKE leads to a change of light polarization in a similar way as the polar MOKE does. The problem here is, however, how to disentangle the out-of-plane magnetization projection, which is sensed by polar MOKE, from the in-plane projection, which is sensed by longitudinal MOKE. To do so, it is sufficient to measure the signal for "$+\theta_i$" and "$-\theta_i$" that can be done by swapping the propagation path of incident and reflected beams [8] or by rotating the sample for 180° about the surface normal [3], which in both cases is, however, connected with rather major changes in the optical setup. Alternatively, the microscope objective with a high numerical aperture oriented along the surface normal can be used [2, 9]. In this case, the necessary large angles of incidence are present at the edges of the laser beams (due to the tight focusment of the light by the objective) and the required parts of the signals are selected by the usage of quadrant detectors [2, 9]. However, as this detection scheme presumes a highly symmetrical beam profile, its application should be done with a caution [2]. Another available MO effect is the transversal MOKE, which is active when magnetization is perpendicular to the light plane of incidence and which is apparent as a change of reflected light intensity [1, 10]. However, as this effect is usually rather weak, the obtained signals are usually noisier than that measured by longitudinal MOKE [10]. A completely different detection scheme, which is based on a Fresnel scattering matrix formalism, was reported very recently [5]. But in this technique a rather complicated calibration procedure has to be performed to obtain the magnetization dynamics from the measured data [5, 11]. The list of the currently existing experimental MO methods can be closed by mentioning the second-harmonic MOKE where information about the in-plane position of magnetization can be deduced from the efficiency of the second harmonic generation (SHG) [4, 12]. However, because SHG is a non-linear optical effect, laser intensities exceeding 10 GW/cm$^2$ have to be used that is usually not compatible with a concept of a weak optical probe.

## STATIC MAGNETO-OPTICAL SIGNAL IN (GA,MN)AS

The magnetization **M** in a material is characterized by its magnitude $M_s$ and orientation, which can be described by the polar angle $\varphi$ and the azimuthal angle $\theta$ - see Fig. 1. In (Ga,Mn)As there are several magneto-optical (MO) effects that can be

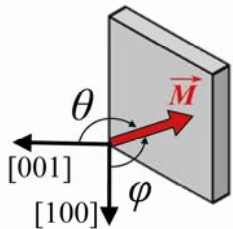

Fig. 1 Definition of the polar angle $\varphi$ and the azimuthal angle $\theta$ that describe the orientation of magnetization in the sample; the angles are counted relative to [100] and [001] crystallographic directions in the sample, respectively.

observed depending on the mutual orientation of the magnetization and the incident light direction [13, 14]. We will limit the discussion to the case when the light beam is close to the normal incidence (in our experiment the angle of incidence is 2° and 8° for pump and probe pulses, respectively). In this geometry only two effects are responsible for the measured MO signal – the polar MOKE and magnetic linear dichroism (MLD). We also note that in the following we will concentrate on the rotation of the polarization plane of the reflected linearly polarized light, but the same applies also for the change of the light ellipticity. In polar

MOKE the rotation of polarization occurs due to the different index of refraction for $\sigma^+$ and $\sigma^-$ circularly polarized light propagating *parallel to the direction of magnetization* - see Fig. 2(a). Consequently, the rotation of light polarization $\Delta\beta$ is proportional to the projection of magnetization to the direction of light propagation

$$MO^{PKE} \equiv \Delta\beta^{PKE} \equiv \beta' - \beta = P^{PKE} \frac{M_z}{M_s} = P^{PKE} \cos\theta_0, \qquad (1)$$

where $\beta$ and $\beta'$ describes the orientation of the input and output linear polarization [see Fig. 3(a)], $P^{PKE}$ is the corresponding magneto-optical coefficient of the sample, $M_s$ and $M_z$ are magnitude and $z$ component of magnetization, and $\theta_0$ describes the equilibrium out-of-plane orientation of magnetization, respectively. Here we adopted the following sign convention: If light is reflected along the direction of magnetization, the value $P^{PKE} > 0$ corresponds to a counterclockwise rotation of incident polarization (i.e., $\Delta\beta > 0$) when viewed by an observer facing the sample – see Fig. 3(a). We note that this MO effect is linear in magnetization (i.e., the sign of $\Delta\beta$ is changed when the direction of magnetization is reversed) and that the value of $\Delta\beta^{PKE}$ does not dependent on $\beta$.

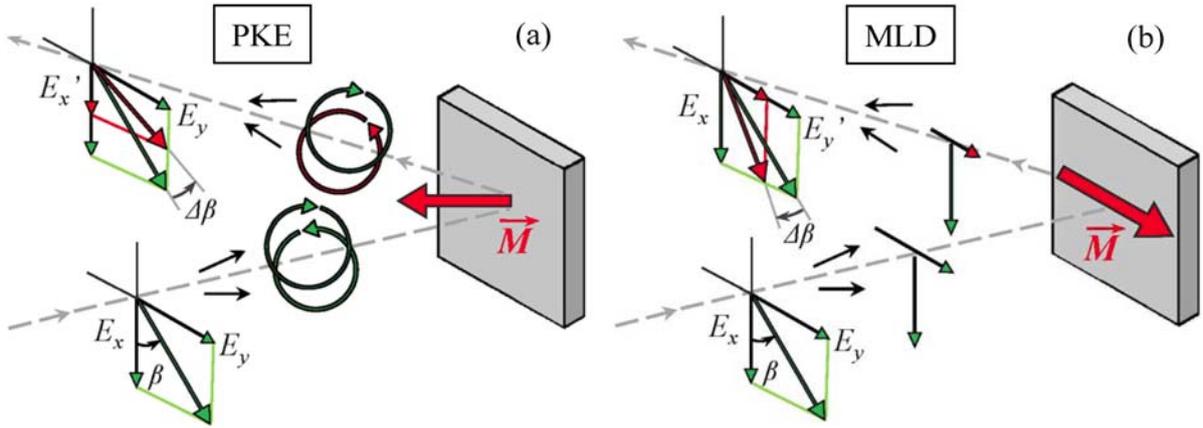

Fig. 2. Schematic illustration of two magneto-optical effects in (Ga,Mn)As that are responsible for a rotation of the polarization plane $\Delta\beta$ of reflected light at normal incidence. (a) Polar Kerr effect (PKE) that is due to the different index of refraction for $\sigma^+$ and $\sigma^-$ circularly polarized light propagating parallel to the direction of magnetization $M$. (b) Magnetic linear dichroism (MLD) that is due to the different absorption (reflection) coefficient for light linearly polarized parallel and perpendicular to $M$ if the light propagates perpendicular to the direction of $M$.

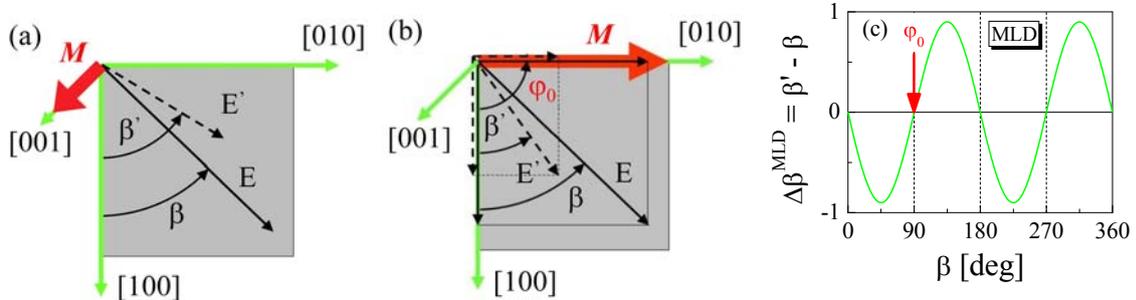

Fig. 3. Polarization dependence of magneto-optical effects. (a) PKE is proportional to the out-of-plane projection of magnetization; for $P^{PKE} > 0$ and $M_z > 0$ this MO effect leads to $\Delta\beta > 0$ for any $\beta$. (b) MLD is sensitive to the in-plane projection of magnetization; the magnitude and sign of $\Delta\beta$ is a harmonic function of $\beta$ as described by Eq. (5) and schematically illustrated in (c) where the vertical red arrow depicts the assumed position of the magnetization.

The second MO effect is the magnetic linear dichroism (MLD) [14], which originates from the different absorption (reflection) coefficient for light linearly polarized parallel and perpendicular to *M*. This effect occurs if the light propagates *perpendicular to the direction of magnetization M* - see Fig. 2(b). To derive the rotation of light polarization due to MLD we first suppose that the magnetization is located in the sample plane with a position characterized by an angle $\varphi_0$ [see Fig. 3(b)]. We can express the projections of the incident electric field amplitude *E* parallel to magnetization ($E^{\parallel}$) and perpendicular to magnetization ($E^{\perp}$) using $\varphi_0$ and the incident polarization orientation $\beta$

$$E^{\parallel} = E\cos(\varphi_0 - \beta), \tag{2a}$$
$$E^{\perp} = E\sin(\varphi_0 - \beta). \tag{2b}$$

The same can be done for the reflected electric field amplitude *E'*. If we now consider that $E^{\parallel}$ ($E^{\perp}$) is reflected from (Ga,Mn)As with the amplitude reflection coefficient *a* (*b*), we obtain

$$tg(\varphi_0 - \beta) = \frac{E^{\perp}}{E^{\parallel}}, \tag{3a}$$

$$tg(\varphi_0 - \beta') = \frac{bE^{\perp}}{aE^{\parallel}}, \tag{3b}$$

from which the rotation of light polarization $\Delta\beta \equiv \beta' - \beta$ can be easily derived

$$tg\Delta\beta = \frac{(a-b)\,tg(\varphi_0 - \beta)}{a + b\,[tg(\varphi_0 - \beta)]^2}. \tag{4}$$

If we now assume that $a/b \approx 1$ (i.e., that $\Delta\beta$ is small) we obtain

$$\Delta\beta = P^{MLD}\sin 2(\varphi_0 - \beta), \tag{5}$$

where the magneto-optical coefficient $P^{MLD}$ is defined as

$$P^{MLD} = 0.5\left(\frac{a}{b} - 1\right). \tag{6}$$

We note that $P^{MLD}$ depends quadratically on the magnetization magnitude $M_s$ [14]

$$P^{MLD} \sim M_s^2. \tag{7}$$

In a more general case, when magnetization has an arbitrary orientation characterized by $\varphi_0$ and $\theta_0$, the rotation of light polarization by MLD is given by

$$MO^{MLD} \equiv \Delta\beta^{MLD} \equiv \beta' - \beta = P^{MLD}\sin\theta_0 \sin 2(\varphi_0 - \beta). \tag{8}$$

The total MO response of any (Ga,Mn)As sample is given by a sum of contributions due to PKE and MLD:

$$MO^{stat} = MO^{PKE} + MO^{MLD} = P^{PKE} \cos\theta_0 + P^{MLD} \sin\theta_0 \sin 2(\varphi_0 - \beta) \qquad (9)$$

The magnitude of $P^{PKE}$ and $P^{MLD}$ can be directly measured if the magnetization is oriented by a strong external magnetic field to the out-of-plane ($\theta_0 = 0°$) and in-plane ($\theta_0 = 90°$; $\varphi_0 - \beta = 45°$) positions, respectively. The measured spectral dependence of $P^{PKE}$ and $P^{MLD}$ for the investigated epilayer are shown in Fig. 4(b).

With no magnetic field applied, the magnetization points to the easy axis direction, which is determined by the magnetic anisotropy of the sample. All the investigated (Ga,Mn)As samples with nominal doping ranging from 1.5% to 14% are in-plane magnets (i.e., $\theta_0 = 90°$)[15]. Consequently, in the equilibrium conditions the static MO signal is only due to MLD

$$MO^{stat} = P^{MLD} \sin 2(\varphi_0 - \beta), \qquad (10)$$

see also Fig. 3(c).

## DYNAMICAL MAGNETO-OPTICAL SIGNAL IN (GA,MN)AS

The impact of a strong pump laser pulse modifies the properties of the sample that leads to the mutual misalignment of the magnetization and the easy axis and, consequently, to the precession of magnetization around the quasi-equilibrium position of the easy axis. The measured dynamical MO signal $\delta MO$, which is a function of the time delay between pump and probe pulses $\Delta t$ and the probe polarization orientation $\beta$, can be fitted well by the phenomenological equation,

$$\delta MO(\Delta t, \beta) = A(\beta) \cos[\omega_{Mn} \Delta t + \Phi(\beta)] e^{-\Delta t/\tau_G} + C(\beta) e^{-\Delta t/\tau_p}, \qquad (11)$$

where $A$ and $C$ are the amplitudes of the oscillatory and pulse function, respectively, $\omega_{Mn}$ is the ferromagnetic moment precession frequency, $\Phi$ is the phase factor, $\tau_G$ is the Gilbert damping time, and $\tau_p$ is the pulse function decay time. The pulse function in $\delta MO$ signal is a transient non-oscillatory change of the static signal $MO^{stat}$. In fact, there are two distinct contributions to this signal. Firstly, there is a contribution due to a change of the quasi-equilibrium magnetization position (the "tilt" signal in the following), which corresponds to a derivative of Eq. (9) with respect to a small change of $\varphi$ and $\theta$. Secondly, the pump-induced demagnetization of the material [16] reduces also the static MO response (the "demagnetization" signal), which is in the investigated samples with the in-plane anisotropy given by Eq. (10). If we assume, for simplicity, that both these signals have the same dynamics, which is presumably dominated by a dissipation of heat from the irradiated spot on the sample, we have the following equation for the measured amplitude of the pulse function $C$ (taking into account that $\theta_0 = 90°$)

$$\boxed{C(\beta) = C_{tilt}(\beta) + C_{demag}(\beta) = -\delta\theta_{qe} P^{PKE} + \delta\varphi_{qe} P^{MLD} 2\cos 2(\varphi_0 - \beta) + \frac{\delta M_s}{M_0} P^{MLD} 2\sin 2(\varphi_0 - \beta)}$$
(12)

where first two terms on the right-hand-side correspond to $C_{tilt}$ and the last term is $C_{demag}$. $\delta\theta_{qe}$ and $\delta\varphi_{qe}$ describe the out-of-plane and in-plane movement of the easy axis (along which the magnetization precesses), respectively, and $\delta M_s / M_0$ characterizes a reduction of the magnetization magnitude (relative to the original value of $M_0$). In fact, Eq. (12) is of fundamental importance for the analysis of the measured MO signals because it enables to determine experimentally if the precession of magnetization is triggered by the out-of-plane or by the in-plane movement of the easy axis. If the out-of-plane movement dominates in $C_{tilt}$, it will not depend on $\beta$. On the other hand, if the in-plane movement dominates in $C_{tilt}$, it will be a harmonic function of $\beta$ [see Eq. (12)]. If both movements are comparable in $C_{tilt}$, it will be again a harmonic function of $\beta$ but in this case there will be an offset. The data shown in Fig. 4(d) clearly illustrates that *the easy axis is tilted in the in-plane direction* in the investigated sample for the used experimental conditions.

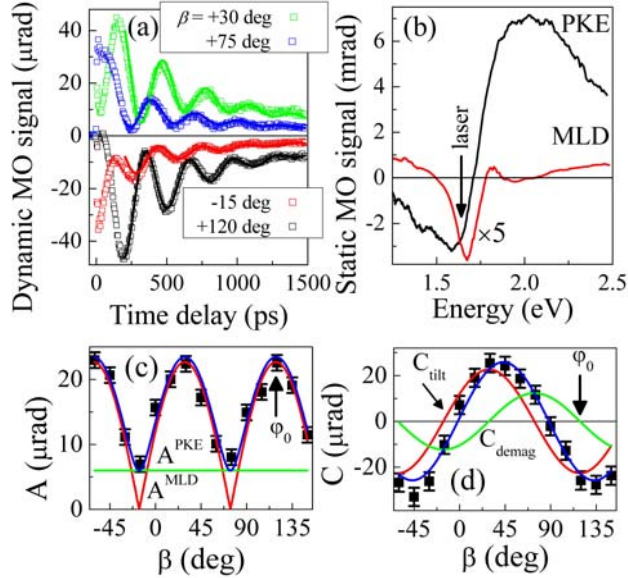

Fig. 4. Magneto-optical signals (polarization rotations) measured in (Ga,Mn)As. (a) Dynamics of the MO signal induced by an impact of pump pulse on the sample that was measured by probe pulses with different probe polarization orientations $\beta$ (points); lines are fits by Eq. (11) with parameters $\omega_{Mn}$ = 20.2 GHz, $\tau_G$ = 360 ps and $\tau_p$ = 1050 ps. (b) Spectral dependence of static PKE and MLD, the arrow indicate the spectral position of the laser pulses used in the time-resolved experiment shown in part (a); note that the data for MLD are multiplied by 5 for clarity. (c) and (d) Polarization dependence of the oscillatory part $A$ (c) and of the pulse function $C$ (d) that was obtained by fitting the dynamics shown in part (a); the values of $A$ and $C$ at time delay of 200 ps are shown (points). Lines are results of simultaneous fits of $A(\beta)$ by Eq. (16) and $C(\beta)$ by Eq. (12). The vertical arrows in (c) and (d) depicts the deduced easy axis position in the sample without the pump pulse, $\varphi_0$. This figure is re-plotted from the main paper (Fig. 2) for convenience.

Also the oscillatory MO signal contains a signal due to the out-of-plane motion of the magnetization, which is sensed by PKE, and a signal due to the in-plane movement of magnetization, which is sensed by MLD. Due to the precessional motion of magnetization these signals are phase shifted for 90° and the total amplitude of the oscillatory MO signal $A$ is given by

$$A(\beta) = \sqrt{\left[A^{MLD}(\beta)\right]^2 + \left[A^{PKE}\right]^2} \ . \tag{13}$$

This equation explains why $A$ depends strongly on the orientation of the probe polarization $\beta$ [see Fig. 4(c)]. The MO signal due to the out-of-plane projection of magnetization (with an amplitude $A^{PKE}$) does not depend on $\beta$ but the MO signal due to the in-plane projection (with an amplitude $A^{MLD}$) is a harmonic function of $\beta$. The position of the maximum in the dependence $A(\beta)$ corresponds to the equilibrium position $\varphi_0$ of the easy axis in the sample (i.e., its position without the pump pulse). This conclusion immediately follows from the fact that the $\beta$ dependence of $A$ comes from the MO signal induced by a change of the in-plane projection of magnetization, which is detected by MLD. And from Fig. 3(c) it is clearly apparent that the strongest change of the MO signal due to an in-plane movement of magnetization is observed when the probe pulses are polarized along the magnetization or perpendicular to it (i.e., when the derivative of Eq. (10) with respect to $\varphi$ is the largest). We recall that prior to this dynamical measurement we oriented the magnetization along the easy axis that is the closest to the [010] crystallographic direction.

The laser pulse-induced shift of the easy axis position is usually much faster than the precessional period $T_{osc} = 2\pi/\omega_{Mn}$ and the Gilbert damping time $\tau_G$. For example, for the data shown in Fig. 4(a) we have $T_{osc}$= 310 ps and $\tau_G$ = 360 ps that is considerably longer than the rise time of the laser-induced transient change of the sample temperature $\Delta T$, which is $\approx$ 30 ps (see Fig. 2 in Ref. 17), and the hole concentration $\Delta p$, which is expected to be quasi-instantaneous. Under these conditions, the initial in-plane amplitude of the oscillations $\delta\varphi$ should be approximately equal to the in-plane movement of the easy axis $\delta\varphi_{qe}$ and, therefore,

$$A^{MLD} \approx C^{shift} = \delta\varphi_{qe} P^{MLD} 2\cos 2(\varphi_0 - \beta). \tag{14}$$

Substituting Eq. (14) to Eq. (13) yields

$$\boxed{A(\beta) = \sqrt{\left[\delta\varphi_{qe} P^{MLD} 2\cos 2(\varphi_0 - \beta)\right]^2 + \left[A^{PKE}\right]^2}.} \tag{16}$$

where $A^{PKE} = -\delta\theta\, P^{PKE}$. Consequently, Eq. (16) and (12) can be used to fit the measured dependences $A(\beta)$ and $C(\beta)$.

The application of this procedure to the measure MO data is shown in Fig. 4(c) and (d). It should be noted that Eq. (14) holds only when the precession of magnetization is fully established – i.e., it is necessary to fit the values of $A$ and $C$ at a time delay that corresponds to the first precessional maximum, which is $\approx$ 200 ps for the case of the data shown in Fig. 4 (see also Fig. 5). As an input to the fitting procedure we used the *independently measured* value of the MO constants for this sample [see Fig. 4 (b)]: $P^{MLD}$ = – 0.59 mrad and $P^{PKE}$ = – 2.65 mrad. As an output we deduced that the easy axis position in the sample without the pump pulse $\varphi_0$ = 119 ± 2°. Moreover, at $\approx$ 200 ps, the in-plane easy axis shift $\delta\varphi_{qe}$ = + 1.1 ± 0.1°, the out-of-plane easy axis shift $\delta\theta_{qe}$ = 0°, and the change of the magnetization magnitude $\delta M_s/M_0$ = – 1.0 ± 0.3%.

## RECONSTRUCTION OF MAGNETIZATION THREE-DIMENSIONAL TRAJECTORY

The obtained detailed understanding of the measured MO signals enables us to perform the full quantitative reconstruction of the real-space trajectory of magnetization from the measured dynamical MO signals without any numerical modeling. Before an impact of the pump pulse the magnetization points to the easy axis direction. From the SQUID measurements we know that the equilibrium easy axis is located in the sample plane, i. e. $\theta_0$ = 90°. From the results shown in Fig. 4(c) we know the in-plane position of the easy axis $\varphi_0$ = 119°. The impact of pump pulse induces a transient increase of the lattice temperature and of the hole concentration that in turn leads to the easy axis shift. Consequently, magnetization starts to follow the easy axis movement by the precessional motion

$$\varphi(\Delta t) = \varphi_0 + \delta\varphi(\Delta t), \quad (17a)$$
$$\theta(\Delta t) = \theta_0 + \delta\theta(\Delta t), \quad (17b)$$

where $\delta\varphi(\Delta t)$ and $\delta\theta(\Delta t)$ and describe the in-plane and out-of-plane transient movement of magnetization, respectively. This movement of magnetization leads to a modification of the static magneto-optical response of the sample ($MO^{stat}$), which is given by Eq. (9). Taking into account that $\theta_0 = 90°$ we have

$$\boxed{\delta MO(\Delta t, \beta) = -\delta\theta(\Delta t)P^{PKE} + \delta\varphi(\Delta t)P^{MLD} 2\cos 2(\varphi_0 - \beta) + \frac{\delta M_s(\Delta t)}{M_0} P^{MLD} 2\sin 2(\varphi_0 - \beta)}.$$
(18)

The first two terms in Eq. (18) are connected with the movement of magnetization and the last term describes the static MO signal change due to the demagnetization. The harmonic dependence of the MO signal due to MLD, which is sensitive to the in-plane motion of magnetization, on the probe polarization orientation $\beta$ enables to separate this signal from the MO signal due to PKE, which is sensitive to the out-of-plane motion of magnetization and which does not depend on $\beta$. It is apparent from Eq. (18) [see also Fig. 4(d)] that for $\beta$ equal to $\varphi_0$ and $\varphi_0 - 90°$ the demagnetization does not contribute to the measured dynamical MO signal. Moreover, contributions to $\delta MO$ due to $\delta\varphi(t)$ are maximal and exactly opposite for $\beta = \varphi_0$ and $\beta = \varphi_0 - 90°$. Consequently,

$$\boxed{\delta\varphi(\Delta t) = [\delta MO(\Delta t, \varphi_0) - \delta MO(\Delta t, \varphi_0 - 90°)]/(4P^{MLD})}. \quad (19)$$

Similarly, for $\beta$ equal to $\varphi_0 - 45°$ and $\varphi_0 - 135°$ the contributions due to $\delta\varphi(t)$ are not present in $\delta MO$ and the signal due to the demagnetization is exactly opposite for these two angles that leads to

$$\boxed{\delta\theta(\Delta t) = -[\delta MO(\Delta t, \varphi_0 - 45°) + \delta MO(\Delta t, \varphi_0 - 135°)]/(2P^{PKE})} \quad (20)$$

$$\boxed{\frac{\delta M_s(\Delta t)}{M_0} = [\delta MO(\Delta t, \varphi_0 - 45°) - \delta MO(\Delta t, \varphi_0 - 135°)]/(4P^{MLD})}. \quad (21)$$

The dynamics of magnetization, which was deduced from the data in Fig. 4(a), is shown in Fig. 5.

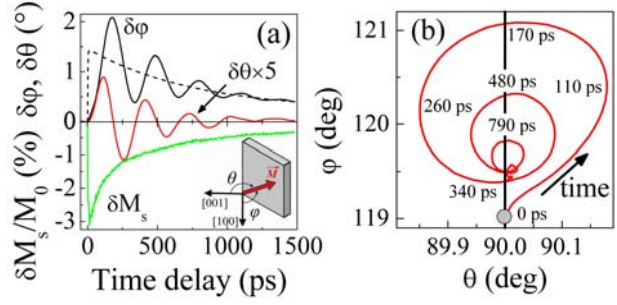

Fig. 5. Reconstruction of the magnetization real-space trajectory. (a) Time evolution of $\delta M_s(t)/M_0$, $\delta\varphi(t)$ and $\delta\theta(t)$; the dotted line depicts the in-plane evolution of the easy axis position around which the magnetization precesses. (b) Orientation of magnetization at different times after the impact of the pump pulse; the sample plane is represented by the vertical line and the equilibrium position of the easy axis is depicted by the grey spot. This figure is re-plotted from the main paper (Fig. 3) for convenience.

## NUMERICAL MODELLING OF DYNAMIC MAGNETO-OPTICAL SIGNAL

## BY LLG EQUATION

To corroborate the model presented above, we performed also a numerical modeling of the measured precessional MO signal by Landau-Lifshitz-Gilbert (LLG) equation. We used LLG equation in spherical coordinates where the time evolution of magnetization magnitude $M_s$ and orientation, which is characterized by the polar $\theta$ and azimuthal $\varphi$ angles, is given by

$$\frac{dM_s}{dt} = 0, \tag{21}$$

$$\frac{d\theta}{dt} = -\frac{\gamma}{(1+\alpha^2)M_s}\left(\alpha \cdot A + \frac{B}{\sin\theta}\right), \tag{22}$$

$$\frac{d\varphi}{dt} = \frac{\gamma}{(1+\alpha^2)M_s \sin\theta}\left(A - \frac{\alpha \cdot B}{\sin\theta}\right), \tag{23}$$

where $\alpha$ is the Gilbert damping parameter and $\gamma$ is the gyromagnetic ratio. Functions $A = dF/d\theta$ and $B = dF/d\varphi$ are the derivatives of the energy density functional $F$ with respect to $\theta$ and $\varphi$, respectively. We expressed $F$ in a form

$$F = M\left[K_c \sin^2\theta\left(\frac{1}{4}\sin^2 2\varphi \sin^2\theta + \cos^2\theta\right) - K_{[001]}\cos^2\theta - \frac{K_{[110]}}{2}\sin^2\theta(1-\sin 2\varphi)\right], \tag{22}$$

where the magnetic anisotropy constants $K_c$, $K_{[001]}$ and $K_{[110]}$ were measured independently by SQUID. To model the measured MO data, we first computed from LLG equation the time-dependent deviations of the spherical angles [$\delta\theta(t)$ and $\delta\varphi(t)$] from the corresponding equilibrium values. Then we calculated how such changes of $\theta$ and $\varphi$ modify the magneto-optical response of the sample [cf. Eq. (18)].

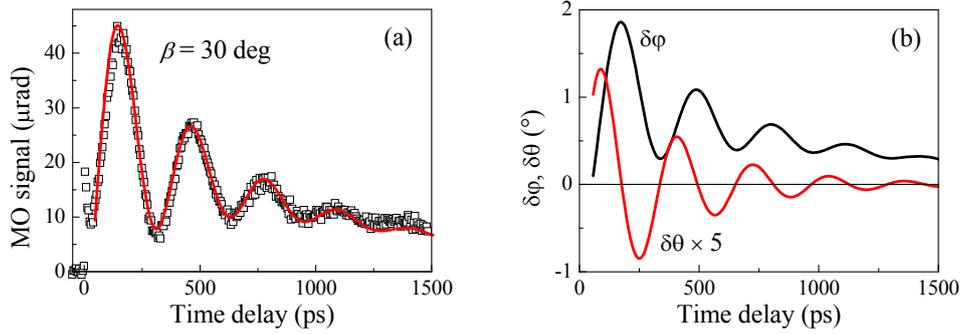

Fig. 6. Numerical modeling of the measured magneto-optical signal by LLG equation. (a) Dynamics of the MO signal measured by probe pulses with $\beta = 30°$ (points); line is the fit by LLG. (b) Calculated time evolutions of $\delta\varphi(t)$ and $\delta\theta(t)$ that were used to model the data shown in (a).

As an example, we show in Fig. 6(a) the results of this numerical model for $\beta = 30°$. The corresponding calculated time evolutions of $\delta\varphi(t)$ and $\delta\theta(t)$, which are depicted in Fig. 6(b), are very similar to that obtained by our direct experimental technique [see Fig. 5(a)]. There are just two small differences. Firstly, the absolute magnitudes of $\delta\varphi(t)$ and $\delta\theta(t)$ are slightly different that is probably connected with the fact that in LLG equation the magnetization magnitude is supposed to be constant [cf. Eq. (21)] that is not exactly fulfilled in our case [see Fig. 5(a)]. Secondly, the experimentally observed rise of $\delta\theta(t)$ [see Fig. 5(a)] is different from that computed by LLG [see Fig. 6(b)]. The most plausible explanation is that in the real experiment the laser-induced heating of the sample has a rise time of about 30 ps while we assumed an abrupt magnetic anisotropy change in our modeling.